# Better Performance ACF Operation for PAPR Reduction of OFDM Signal

Md. Munjure Mowla[a], Md. Yeakub Ali[b] and Abdulla Al Suman[c]

[a,b,c] Department of Electronics & Telecommunication Engineering,
Rajshahi University of Engineering & Technology, Rajshahi, Bangladesh
[a]rimonece@gmail.com ; [b]yeakub.ruet08@gmail.com2 ; [c]suman.ete.ruet@gmail.com

**Abstract.** Orthogonal frequency division multiplexing (OFDM) is a promising modulation radio access scheme for next generation wireless communication systems because of its inherent immunity to multipath interference due to a low symbol rate, the use of a cyclic prefix, and its affinity to different transmission bandwidth arrangements. OFDM has already been adopted as a radio access scheme for several of the latest cellular system specifications such as the long-term evolution (LTE) system in the 3GPP (3rd Generation Partnership Project). Nevertheless, peak-to-average power ratio (PAPR) of OFDM signal is a significant drawback since it restricts the efficiency of the transmitter. A number of promising approaches have been proposed & implemented to reduce PAPR with the expense of increase transmit signal power, bit error rate (BER) & computational complexity and data rate loss, etc. In this paper, a relatively better scheme of amplitude clipping & filtering operation (ACF) is proposed and implemented which shows the significant improvement in case of PAPR reduction while increasing slight BER compare to an present method.

**Keywords:** Bit Error rate (BER), Complementary Cumulative Distribution Function (CCDF), Clipping Ratio (CR), Orthogonal Frequency Division Multiplexing (OFDM) and Peak to Average Power Ratio (PAPR).

## 1 INTRODUCTION

During the last few years, the wireless communication industries has gone through several improvement stages in the very fastest way and as a result of that the demand on the wireless services has growth rapidly as well. Due to the unpredictable nature of the wireless channels, calculating the propagation and the noise is not easy. On the other hand, the wired channels have less complexity of calculating the noise because that the signal propagates in a fixed path. One of the reasons to have signal degradation is the Additive White Gaussian Noise (AWGN) that could be occurred due to industrial or natural sources. It is much easier to use single-carrier transmission scheme due to the simplicity and accuracy that is provided especially with the low data rate. This technique has its own advantages like the simplicity of transmitting the signal through a flat fading channel and saving more power since there is no need to extend the bandwidth by inserting guard interval (Albdran, S. 2012). However, the use of single-carrier may have actual drawbacks with high data rate including equalizing complexity. OFDM is used to overcome the shortages of the single-carrier transmission scheme in the case of having high data rate. The high bandwidth efficiency is one of the OFDM advantage in the case of having a big number of subcarriers. Also, with the OFDM scheme there is much lower chance to have ISI.

One of the major drawbacks of OFDM is the high peak-to-average power ratio (PAPR) of the transmit signal. If the peak transmit power is limited by either regulatory or application constraints, the effect is to reduce the average power allowed under multicarrier transmission relative to that under constant power modulation techniques. This in turn reduces the range of multicarrier transmission. In many low-cost applications, the drawback of high PAPR may outweigh all the potential benefits of multicarrier transmission systems (Han, S. H. 2005). A number of promising approaches or processes have been proposed & implemented to reduce PAPR with the expense of increase transmit signal power, bit error rate (BER) & computational complexity and loss of data rate, etc. So, a system trade-off is required between these. These techniques include Amplitude Clipping and Filtering, Peak Windowing, Peak Cancellation, Peak Reduction Carrier, Partial Transmit Sequence (PTS), Selective Mapping (SLM), Tone Reservation (TR), Tone Injection (TI) etc (Mowla, M. M. 2013).

In this paper, amplitude clipping and filtering operation (ACF) is applied to reduce the PAPR with a relatively new filtering design which reduces the PAPR considerably at the expense of BER slightly. Clipping is done in the time domain and after that the clipped signal is passed through the filter which is also composed of FFT, IFFT & band pass filter.

## 2 BASICS OF OFDM SYSTEM AND PAPR

In this section, we discuss about the basics format of OFDM systems and the fundamental discussion of PAPR & the motivation of reducing PAPR.

### 2.1 OFDM Systems

In OFDM systems, a fixed number of successive input data samples are modulated first (e.g, PSK or QAM), and then jointly correlated together using inverse fast Fourier transform (IFFT) at the transmitter side. IFFT is used to produce orthogonal data subcarriers. Let, data block of length $N$ is represented by a vector, $X=[X_0, X_1......X_{N-1}]^T$. Duration of any symbol $X_K$ in the set $X$ is $T$ and represents one of the sub-carriers set. As the N sub-carriers chosen to transmit the signal are orthogonal, so we can have, $f_n = n\Delta f$, where $n\Delta f = 1/NT$ and $NT$ is the duration of the OFDM data block $X$. The complex data block for the OFDM signal to be transmitted is given by (Han, S. H. 2005),

$$x(t) = \frac{1}{\sqrt{N}} \sum_{n=0}^{N-1} X_n e^{j2\pi n \Delta f t} \qquad 0 \leq t \leq NT \qquad (1)$$

Where,

$j=\sqrt{-1}$, $\Delta f$ is the subcarrier spacing and $NT$ denotes the useful data block period.

### 2.2 PAPR

Presence of large number of independently modulated sub-carriers in an OFDM system the peak value of the system can be very high as compared to the average of the whole system shown in fig. 1. Coherent addition of N signals of same phase produces a peak which is N times the average signal (Jiang, T. 2008).

The PAPR of the transmitted signal is defined as (Revuelto, N. 2008),

$$PAPR[x(t)] = \frac{\max_{0 \leq t \leq NT} |x(t)|^2}{P_{av}} = \frac{\max_{0 \leq t \leq NT} |x(t)|^2}{\frac{1}{NT} \int_0^{NT} |x(t)|^2 \, dt} \qquad (2)$$

Where, Pav is the average power of and it can be computed in the frequency domain because Inverse Fast Fourier Transform (IFFT) is a (scaled) unitary transformation.

### 3 CONVENTIONAL CLIPPING AND FILTERING

Amplitude Clipping and Filtering is considered as the simplest technique which may be under taken for PAPR reduction in an OFDM system. A threshold value of the amplitude is set in this case to limit the peak envelope of the input signal (Revuelto, N. 2008).

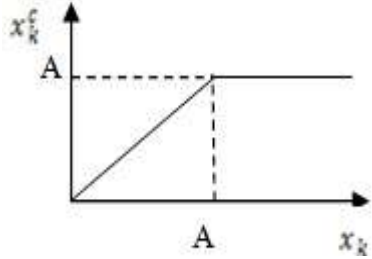

Fig. 1. Clipping Function

The clipping ratio (CR) is defined as,

$$CR = \frac{A}{\sigma} \qquad (3)$$

Where, A is the amplitude and $\sigma$ is the root mean squared value of the unclipped OFDM signal.

The clipping function shown in fig. 1 is performed in digital time domain, before the D/A conversion and the process is described by the following expression,

$$x_k^c = \begin{cases} x_k & |x_k| \leq A \\ Ae^{j\phi(x_k)} & |x_k| > A \end{cases} \qquad 0 \leq k \leq N-1 \qquad (4)$$

Where, $x_k^c$ is the clipped signal, $x_k$ is the transmitted signal, A is the amplitude and $\phi(x_k)$ is the phase of the transmitted signal $x_k$.

### 4. PROPOSED CLIPPING AND FILTERING SCHEME

Mentioning the third limitation in Reference (Han, S. H. 2005), that is clipped signal passed through the BPF causes less BER degradation, we design a new scheme for clipping & filtering method where clipped signal will pass through the band pass filter (BPF). This proposed scheme is shown in the fig. 2. It shows a block diagram of a PAPR reduction scheme using clipping and filtering, where *L* is the oversampling factor and *N* is the number of subcarriers. The input of the IFFT block is the interpolated signal introducing *N(L −1)* zeros (being *L* the oversampling rate) (also, known as zero padding) in the middle of the original signal is expressed as (Cho,Y.S. 2010),

$$X'[k] = \begin{cases} X[k], & \text{for } 0 \leq k \leq \frac{N}{2} \text{ and } NL - \frac{N}{2} < k < NL \\ 0 & \text{elsewhere} \end{cases} \qquad (5)$$

In this scheme, the L-times oversampled discrete-time signal is generated as,

$$x'[m] = \frac{1}{\sqrt{L.N}} \sum_{k=0}^{L.N-1} X'[k].e^{\frac{j2\pi m \Delta f k}{L.N}}, \qquad m = 0,1,...NL - 1 \qquad (6)$$

and is then modulated with carrier frequency fc to yield a passband signal $x^p[m]$.

Let $x_c^p[m]$ denote the clipped version of $x^p[m]$, which is expressed as,

$$x_c^p[m] = \begin{cases} -A & x^p[m] \leq -A \\ x^p[m] & |x^p[m]| < A \\ A & x^p[m] \geq A \end{cases} \quad (7)$$

Where, A is the pre-specified clipping level. Eqn.( 7) can be applied only to the passband signals.

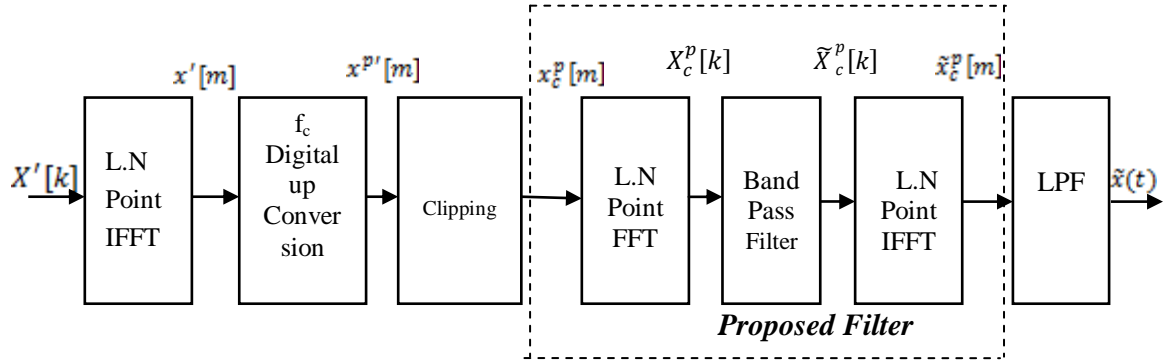

Fig. 2. Block Diagram of Proposed Clipping & Filtering Scheme.

After clipping, the signals are passed through the filters. The filter itself consists on a set of FFT-IFFT operations where filtering takes place in frequency domain after the FFT function.

The FFT function transforms the clipped signal $x_c^p[m]$ to frequency domain yielding $X_c^p[k]$. The information components of $X_c^p[k]$ are passed to a band pass filter (BPF) producing $\tilde{X}_c^p[k]$. This filtered signal is passed to the unchanged condition of IFFT block and the out-of-band radiation that fell in the zeros is set back to zero. The IFFT block of the filter transforms the signal to time domain and thus obtain $\tilde{x}_c^p[m]$.

## 5. DESIGN AND SIMULATION PARAMETERS

In this simulation, an IIR filter using the Chebyshev Type I method is used in the composed filtering. Table 1 shows the values of parameters used in the QPSK system for analyzing the performance of clipping and filtering technique (Cho,Y.S. 2010). We have simulated the both scheme in the same parameters at first.

Table 1. Parameters Used for Simulation of Clipping and Filtering

| Parameters | Value |
|---|---|
| Bandwidth ( BW) | 1 MHz |
| Over sampling factor (L) | 8 |
| Sampling frequency, $f_s$ = BW*L | 8 MHz |
| Carrier frequency, $f_c$ | 2 MHz |
| FFT Size / No. of Subscribers (N) | 128 |
| Modulation Format | QPSK |
| Clipping Ratio (CR) | 0.8, 1.0, 1.2, 1.4, 1.6 |

Using the special type of bandpass filter in the composed filter, significant improvement is observed in the case of PAPR reduction. But, when the clipped & filtered signal is passed through the AWGN channel, the BER is slightly increased compare to the existing method. Chebyshev Type I filters are equiripple in the passband and monotonic in the stopband. Type I filters roll off faster than type II filters, but at the expense of greater deviation from unity in the passband. The filter is optimal in the sense that the maximum error between the desired frequency response and the actual frequency response is minimized.

### 5.1 Simulated Results for PAPR Reduction

In this paper, simulation is performed for different CR values and compare with an existing method (Cho,Y.S. 2010). At first, we simulate the PAPR distribution in the existing algorithm for CR values = 0.8, 1.0, 1.2, 1.4, 1.6 with QPSK modulation and N=128.

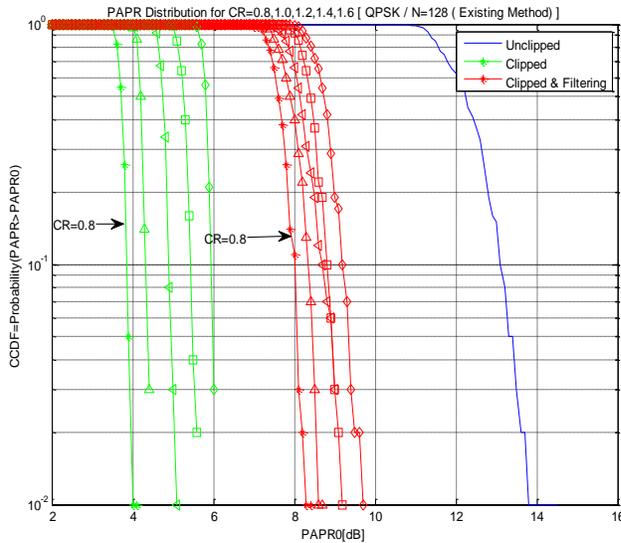 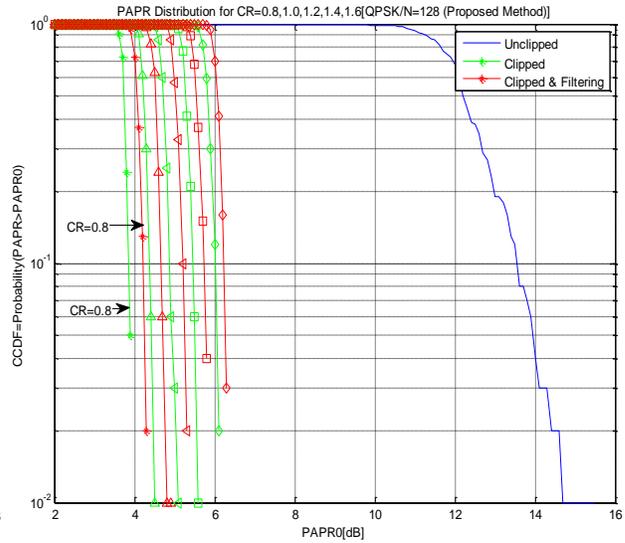

Fig. 3. PAPR Distribution for CR=0.8,1.0,1.2,1.4,1.6 [QPSK / N=128 (Existing)]

Fig. 4. PAPR Distribution for CR=0.8,1.0,1.2,1.4,1.6 [QPSK / N=128 (Proposed)]

In the Proposed method, simulation shows the significant reduction of PAPR which is shown for different CR values in fig. 4. At CCDF $=10^{-1}$, the unclipped signal value is 13.51 dB and others values for different CR are tabulated in the table 2.

Table 2: Comparison of Existing with Proposed Method for PAPR value [QPSK / N=128]

| CR value | PAPR value (dB) (Existing) | PAPR value (dB) (Proposed) | Improvement in PAPR value (dB) |
| --- | --- | --- | --- |
| 0.8 | 8.10 | 4.21 | 3.89 |
| 1.0 | 8.35 | 4.67 | 3.68 |
| 1.2 | 8.72 | 5.21 | 3.51 |
| 1.4 | 8.81 | 5.72 | 3.09 |
| 1.6 | 9.22 | 6.23 | 2.99 |

From table 2, it is clearly observed that the proposed method reduces PAPR significantly with respect to existing (Cho,Y.S. 2010) analysis. For the lower value of CR, PAPR improvement is higher but with the increasing CR value, improvement rate is decreasing.

### 5.2 Simulated Results for BER Characteristics

The clipped & filtered signals for different CR values are passed through the AWGN channel and BER are measured for both existing & proposed methods. We have also simulated the analytical BER that is shown in the curve. It can be seen from these figures that the BER performance becomes worse as the CR decreases. That means, for low value of CR, the BER is more. At first, existing method is simulated for QPSK & N=128 for the same data mentioned in table 1 and resulted graph is shown in fig. 5. The measured BER values at 6 dB point are tabulated in table 3.

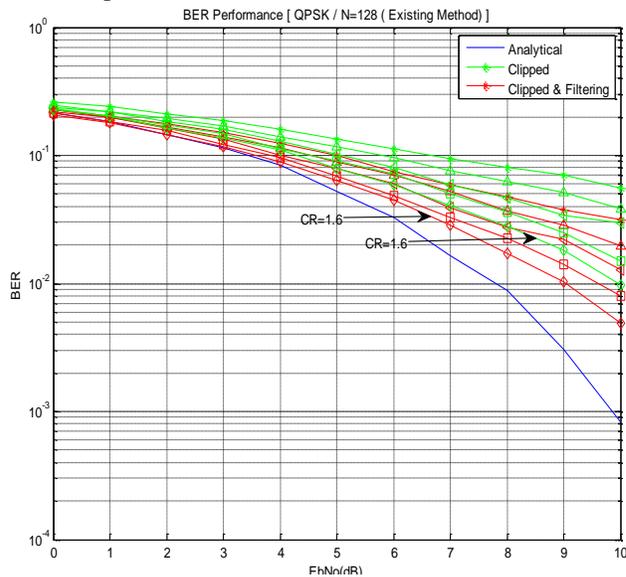 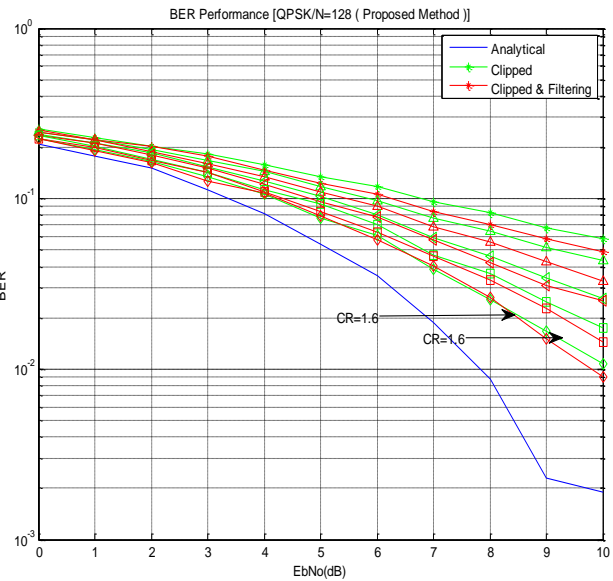

Fig. 5.  BER Performance [QPSK / N=128 (Existing)]    Fig. 6.  BER Performance [QPSK / N=128 (Proposed)]

Simulation is executed for proposed method using same parameters and observed that BER increases very slightly with respect to existing method for same value of CR.  Fig. 6 shows the BER performance when both clipping and clipped & filtering techniques are used.

Table 3: Comparison of Existing with Proposed Method for BER value [QPSK / N=128]

| CR value | BER value (Existing) | BER value (Proposed) | Difference in BER value |
|---|---|---|---|
| 0.8 | 0.07413 | 0.10631 | -0.03218 |
| 1.0 | 0.06984 | 0.09012 | -0.02028 |
| 1.2 | 0.05982 | 0.07846 | -0.01864 |
| 1.4 | 0.04891 | 0.06358 | -0.01467 |
| 1.6 | 0.04449 | 0.05748 | -0.01299 |

From fig. 5 & fig. 6, it is observed that BER performance is slightly worse in proposed method than existing method for different value of CR. The measured BER values at 6 dB point are tabulated in table 3.

From table 3, it is observed that, for CR values (0.8,1.0,1.2,1.4 & 1.6) , the difference magnitude between existing & proposed method are 0.03218,0.02028,0.01864,0.01467 & 0.01299 respectively. These BER degradations are acceptable as these are very low values. It is also investigated that for large number of subscribers (N>128), existing method does not work properly. But, the proposed method works well.

**6. CONCLUSIONS**

In this paper, a comparatively different scheme of ACF operation is performed to reduce PAPR has been analyzed where PAPR reduces significantly compare to an existing method with slightly increase of BER. At first stage, simulation has been executed for existing method with QPSK modulation and number of subscriber (N=128) and then executed for the proposed method for same parameter and observed that PAPR reduces significantly. In the next stage, BER performance is measured for AWGN channel and observed slightly degraded result in the proposed method. No specific PAPR reduction technique is the best solution for multicarrier transmission systems. In the present simulation study, ideal channel characteristics have been assumed. In order to evaluate the OFDM system performance in real world, Rayleigh fading channel will be considerate in future. The increase number of subscribers (N) & other modulation format like QAM or higher order PSK might be another assumption for further study.

**References**

Albdran, S., Alshammari, A. & Matin, M. (2012). Clipping and Filtering Technique for reducing PAPR in OFDM. *IOSR Journal of Engineering (IOSRJEN)*, vol. 2, no.9,pp.91-97.

Cho,Y.S., Kim. J., Yang, W.Y. & Kang,C.G. (2010). *MIMO OFDM Wireless Communications with MATLAB*, Singapore: John Wiley & Sons.

Han, S. H., & Lee, J.H. (2005). An Overview of Peak-To-Average Power Ratio Reduction Techniques For Multicarrier Transmission. *IEEE Wireless Comm,* vol. 12, no.2, pp.56-65.

Jiang, T. & Wu, Y. (2008). An Overview: Peak-To-Average Power Ratio Reduction Techniques for OFDM Signals. *IEEE Transactions On Broadcasting,* vol. 54, no. 2, pp. 257-268.

Mowla, M. M., Razzak, S. M A. & Goni, M. O. (2013). *Improvement of PAPR Reduction for OFDM Signal in LTE System*. Germany: LAP LAMBERT Academic Publishing.

Revuelto, N. (2008). PAPR reduction in OFDM systems. *M.Sc Dissertation*, Universitat Politecnica de Catalunya, Spain.